\documentclass[11pt,a4paper]{article}
\usepackage{jcappub}
\usepackage{hyperref}
\usepackage{epsfig}
\usepackage{color}

\usepackage{amssymb}

\title{Interpretation of the Depths of Maximum  of 
Extensive Air Showers Measured by the Pierre Auger Observatory
}

\collaboration{The Pierre Auger Collaboration}
\author{
\par\noindent
P.~Abreu$^{61}$, 
M.~Aglietta$^{49}$, 
M.~Ahlers$^{90}$, 
E.J.~Ahn$^{78}$, 
I.F.M.~Albuquerque$^{15}$, 
I.~Allekotte$^{1}$, 
J.~Allen$^{82}$, 
P.~Allison$^{84}$, 
A.~Almela$^{11,\: 7}$, 
J.~Alvarez Castillo$^{54}$
% ...
}
\affiliation{
\par\noindent
$^{1}$ Centro At\'{o}mico Bariloche and Instituto Balseiro (CNEA-UNCuyo-CONICET), San 
Carlos de Bariloche, 
Argentina \\
% ...
$^{7}$ Instituto de Tecnolog\'{\i}as en Detecci\'{o}n y Astropart\'{\i}culas (CNEA, CONICET, UNSAM), 
Buenos Aires, 
Argentina \\
% ...
$^{11}$ Universidad Tecnol\'{o}gica Nacional - Facultad Regional Buenos Aires, Buenos Aires,
Argentina \\
% ...
$^{15}$ Universidade de S\~{a}o Paulo, Instituto de F\'{\i}sica, S\~{a}o Paulo, SP, 
Brazil \\
% ...
$^{49}$ Istituto di Fisica dello Spazio Interplanetario (INAF), Universit\`{a} di Torino and 
Sezione INFN, Torino, 
Italy \\
% ...
$^{54}$ Universidad Nacional Autonoma de Mexico, Mexico, D.F., 
Mexico \\
% ...
$^{61}$ LIP and Instituto Superior T\'{e}cnico, Technical University of Lisbon, 
Portugal \\
% ...
$^{78}$ Fermilab, Batavia, IL, 
USA \\
% ...
$^{82}$ New York University, New York, NY, 
USA \\
% ...
$^{84}$ Ohio State University, Columbus, OH, 
USA \\
% ...
$^{90}$ University of Wisconsin, Madison, WI, 
USA \\
% ...
\par\noindent
(\ddag) Deceased \\
(a) Now at Konan University \\
(b) Also at the Universidad Autonoma de Chiapas on leave of absence from Cinvestav \\
(c) Now at University of Maryland \\
(d) Now at NYU Abu Dhabi \\
}
\emailAdd{auger\_spokespersons@fnal.gov}

\abstract{
To interpret the mean depth of cosmic ray air shower maximum and its 
dispersion, 
we parametrize those two observables as functions of the first two moments 
of the $\ln A$ distribution.  We examine the goodness of this simple method 
through simulations of test mass distributions.  The application 
of the parameterization to Pierre Auger Observatory data allows one to study 
the energy dependence of the mean $\ln A$ and of its variance 
under the assumption of selected hadronic interaction models.  We discuss 
possible implications of these dependences in term of interaction 
models and astrophysical cosmic ray sources.}


\begin{document}

\flushbottom
%\linenumbers\relax
\maketitle

\section{Introduction  \label{sec.Intro}}
The most commonly used shower observables for the study of  the composition of Ultra
High Energy Cosmic Rays (UHECR) are the mean value of the depth 
of shower maximum, $\langle X_\mathrm{max} \rangle$, and its dispersion, $\sigma (X_\mathrm{max})$. 
Inferring the mass composition from these measurements is subject to some level 
of uncertainty. This is 
because their conversion to mass relies on
 the use of shower simulation codes which include the assumption of a hadronic
interaction model. The various interaction models~\cite{revHadInt}  have in common the ability to 
fit lower energy accelerator data. However, different physical 
assumptions are used to extrapolate these low energy interaction properties to higher energies. 
Consequently 
they provide different expectations for  $\langle X_\mathrm{max} \rangle$ and  $\sigma (X_\mathrm{max})$.
The first aim of this paper is to discuss how the mean value of the depth of 
shower maximum and its dispersion can be used to interpret mass composition 
even in the presence of uncertainties in the hadronic interaction modeling.

Furthermore, we discuss the different roles of the two observables, 
$\langle X_\mathrm{max} \rangle$ and $\sigma (X_\mathrm{max})$, with respect to mass
composition.
In the interpretation of data they are often used 
 as different, and independent, aspects of the same 
phenomenon. However it is not true to say that both parameters reflect
the cosmic ray composition to the same extent. 
According to
the superposition model~\cite{Gaisser} $\langle X_\mathrm{max} \rangle$ is 
linear in $\langle \ln A \rangle$ and therefore it actually measures mass 
composition
for both pure and mixed compositions. But, we will show that the behaviour 
of $\sigma (X_\mathrm{max})$ is more complex to interpret as there is no one-to-one
 correspondence between its value and a given mean log mass. 
Only in the case of pure composition is this correspondence unique.

In this paper we refine the analysis method originally proposed 
by Linsley~\cite{Linsley83,Linsley85} 
and apply it to the Auger data.
The Pierre Auger Collaboration has published results on the mean and 
dispersion of the $X_\mathrm{max}$ distribution at energies above  
$10^{18}$ eV~\cite{Auger,AugerICRC2011}. 
In this work we apply the proposed method to convert those observables to
the first moments of the log mass distribution, namely $\langle \ln A \rangle$
and $\sigma^2_{{\ln} A}$.

The paper is organized as follows. In Section \ref{sec.Stat} we discuss the 
parameterization for $\langle X_\mathrm{max} \rangle$ 
and $\sigma (X_\mathrm{max})$. 
In Sec.  \ref{sec.Valid}  we test the method with shower
simulations assuming different mass distributions. Sec. \ref{sec.Data} 
describes the application of the method to data. The discussion of
the results and the
conclusions follow in sections \ref{sec.discussion} and \ref{sec.Conc} respectively. The details of the parameterization and the best fit values
for the hadronic interaction models are summarized in Appendix \ref{sec.Pars}.

\section{A method to interpret $\langle X_\mathrm{max} \rangle$ and $\sigma(X_\mathrm{max})$  \label{sec.Stat}}
The interpretation of  $\langle X_\mathrm{max} \rangle$ and $\sigma(X_\mathrm{max})$ can be simplified  by
making use of an analysis method based on  
the generalized Heitler model of extensive air showers~\cite{JMHeitler}. 
In this context $\langle X_\mathrm{max} \rangle$ is a linear function of the 
logarithm of the shower energy per nucleon: 
\begin{equation}
\langle X_\mathrm{max} \rangle = X_{0} +  D~{\log_{10}} \left( \frac{E}{E_0A} \right)~,
\label{eq:XmaxS}
\end{equation}
where $X_{0}$ is the mean depth of proton showers at energy $E_0$ and 
$D$ is the elongation rate~\cite{Elong1,Elong2,Elong3}, i.e., the change of
$\langle X_\mathrm{max} \rangle$  per decade of energy. 
The High Energy hadronic interaction models used in this work are
EPOS 1.99~\cite{epos}, Sibyll~2.1~\cite{Sibyll}, QGSJet~01~\cite{qgsjet01}
and QGSJet~II~\cite{qgsjetII}. Simulated data show that eq. (\ref{eq:XmaxS}) 
gives a fair description 
of EPOS and Sibyll results in the full range of interest for this work, 
10$^{18}$ to 10$^{20}$~eV, but does not reproduce accurately QGSJet models. 
For this reason we generalize the original representation as:
\begin{equation}
\langle X_\mathrm{max} \rangle = X_{0} +  D~{\log_{10}} \left( \frac{E}{E_0A} \right)
+ \xi \ln A + \delta ~\ln A ~{\log_{10}} \left( \frac{E}{E_0} \right)~,
\label{eq:XmaxG}
\end{equation}
where the parameters $\xi$ and $\delta$ are expected to be zero if the model 
predictions are compatible with the superposition result~(\ref{eq:XmaxS}).

For nuclei of the same mass $A$ one expects the shower maximum to be on average:
\begin{equation}
\langle X_\mathrm{max} \rangle = \langle X_\mathrm{max} \rangle_p +  f_E~\ln A ~,
\label{eq:Xmax}
\end{equation}
and its dispersion to be only influenced by shower-to-shower fluctuations:
\begin{equation}
\sigma^2(X_\mathrm{max}) = \sigma^2_\mathrm{sh}(\ln A)~.
\label{eq:varPure}
\end{equation}
Here $\langle X_\mathrm{max} \rangle_p$ denotes the mean depth at maximum 
of proton showers, as obtained 
from either eq. (\ref{eq:XmaxS}) or   (\ref{eq:XmaxG}), and
$\sigma^2_\mathrm{sh}(\ln A)$ is the $X_\mathrm{max}$ variance for mass 
$A$, 
$\sigma^2_\mathrm{sh}(\ln A)=\sigma^2 (X_\mathrm{max} | \ln A)$.
The energy dependent parameter $f_E$ appearing in (\ref{eq:Xmax}) is:
\begin{equation}
f_E = \xi - \frac{D}{\ln 10} + \delta ~{\log_{10}} \left( \frac{E}{E_0} \right).
\label{eq:fE}
\end{equation}
The values of the parameters $X_0$, $D$, $\xi$, $\delta$ depend on the specific
hadronic interaction model.  In this work they are 
obtained from CONEX~\cite{conex} shower simulations as described in 
Appendix~\ref{sec.Pars}.

In the case of a mixed composition at the top of the atmosphere, 
the  mean and variance of $X_\mathrm{max}$ depend on 
the ${\ln}A$ distribution. There are two independent sources of fluctuations: 
the intrinsic shower-to-shower 
fluctuations and the ${\ln}A$ dispersion
arising from the mass distribution. The first term gives rise to 
$\langle \sigma^2_\mathrm{sh} \rangle$, the average variance  of $X_\mathrm{max}$ weighted 
according to the  ${\ln}A$ distribution. 
The second contribution can be written as
$\left(\frac{d \langle X_\mathrm{max} \rangle}{d {\ln}A} \right)^2 \sigma^2_{{\ln} A}$
where $\sigma^2_{{\ln A}}$ is the variance of the $\ln A$ distribution.
We can finally write for the two profile observables:
\begin{equation}
\langle X_\mathrm{max} \rangle = \langle X_\mathrm{max}\rangle_p +  f_E~\langle \ln A \rangle 
\label{eq:modXmax}
\end{equation}
\begin{equation}
\sigma^2(X_\mathrm{max}) =  \langle \sigma^2_\mathrm{sh} \rangle + f_E^2 ~ \sigma^2_{{\ln} A}~.
\label{eq:modSXmax}
\end{equation}
The two equations depend on energy through the parameters but also
via $\langle \sigma^2_\mathrm{sh} \rangle$ and the possible dependence of 
the two moments of the $\ln A$ distribution. 

To obtain an explicit expression for $\langle \sigma^2_\mathrm{sh} \rangle$ we
need a parameterization for $\sigma^2_\mathrm{sh}(\ln A)$. 
We assume a quadratic law in $\ln A$:
\begin{equation}
\sigma^2_\mathrm{sh}(\ln A) = \sigma^2_p [1 + a \ln A + b (\ln A)^2]~,
\label{eq:sigsh}
\end{equation} 
where $\sigma^2_p$ is the $X_\mathrm{max}$ variance for proton showers. 
The evolution of $\sigma^2_\mathrm{sh}(\ln A)$ with energy is 
included in $\sigma^2_p$ and the parameter $a$:
\begin{equation}
\sigma^2_p = p_0 + p_1 \log_{10} \left( \frac{E}{E_0} \right) + p_2 
\left[ \log_{10} \left( \frac{E}{E_0} \right) \right]^2 \quad \mathrm{and} \quad
a = a_0 + a_1 \log_{10} \left( \frac{E}{E_0} \right).
\label{eq:sigshE}
\end{equation} 
The parameters $p_0$, $p_1$, $p_2$, $a_0$, $a_1$, $b$  depend on hadronic 
interactions: the values used in the paper are given in Appendix~\ref{sec.Pars}.

Using measurements of $\langle X_\mathrm{max}\rangle$  and $\sigma (X_\mathrm{max})$, 
equations (\ref{eq:modXmax}) and (\ref{eq:modSXmax}) can be inverted to
get the first two moments of the $\ln A$ distribution. 
From eq. (\ref{eq:modXmax}) one gets:
\begin{equation}
\langle \ln A \rangle = \frac{\langle X_\mathrm{max} \rangle - \langle X_\mathrm{max}\rangle_p}{f_E}~.
\label{eq:meanlnAinv}
\end{equation} 
Averaging eq. (\ref{eq:sigsh}) on $\ln A$ one obtains:
\begin{equation}
\langle \sigma^2_\mathrm{sh} \rangle = \sigma^2_p [1 + a \langle \ln A \rangle + b \langle (\ln A)^2 \rangle]~.
\label{eq:avsigsh}
\end{equation} 
Substituting in eq. (\ref{eq:modSXmax}) we get:
\begin{equation}
\sigma^2(X_\mathrm{max}) =  \sigma^2_p [1 + a \langle \ln A \rangle + 
b \langle (\ln A)^2 \rangle] + f_E^2  \sigma^2_{{\ln} A} ~.
\label{eq:rewrSXmax}
\end{equation} 
But by definition $\langle (\ln A)^2 \rangle = \sigma^2_{{\ln} A} + 
\langle \ln A \rangle^2$. Solving in $\sigma^2_{{\ln} A}$ one finally
obtains:
\begin{equation}
\sigma^2_{\ln A} = \frac{\sigma^2(X_\mathrm{max}) - \sigma^2_\mathrm{sh}(\langle \ln A \rangle)}
 {b ~\sigma^2_p  + f_E^2}~.
\label{eq:VarlnAinv}
\end{equation} 
Equations (\ref{eq:meanlnAinv}) and  (\ref{eq:VarlnAinv}) are the key tools 
used throughout this work for interpreting Pierre Auger Observatory data in terms 
of mass composition and assessing the validity of available hadronic 
interaction models.

\section{Testing the method with simulation \label{sec.Valid}}
Equations (\ref{eq:modXmax}) and (\ref{eq:modSXmax}) can be tested with 
simulations. They contain parameters depending on the hadronic interaction properties and  on 
the mass distribution of nuclei. 
The mass distribution of nuclei refers 
to those nuclei hitting the Earth's atmosphere: it does 
not matter what is the source of the mass dispersion, either a mixed 
composition at injection  or the dispersion caused by propagation. So,
in order to test the method we will simply use different test distributions of the masses at the
top of the atmosphere.

For this purpose we have chosen three different mass distributions:
\begin{enumerate}\addtolength{\itemsep}{-0.02\baselineskip}
\item A distribution uniform in $\ln A$ from $\ln (1)$ to $\ln (56)$ and 
independent 
of energy. The values of $\langle \ln A \rangle$ and $\sigma_{{\ln} A}$ are
respectively 2.01 and 1.16.
\item A Gaussian $\ln A$ distribution with $\langle \ln A \rangle$ 
increasing linearly with $\log E$ from $\ln(4)$ at 
10$^{18}$ eV to $\ln(14)$ at 
10$^{20}$ eV and $\sigma_{{\ln} A} = 0.75$ independent of energy. The Gaussian
is truncated to less than 2 sigmas to avoid unphysical mass values. In this 
case the ${\ln} A$ dispersion is fixed and equal to 0.66 but 
$\langle \ln A \rangle$ varies with energy.
\item Two masses, H and Fe, with proton fraction $H/(H+Fe)$ decreasing  
linearly with $\log E$ from 1 at 10$^{18}$ eV to 0 at 10$^{20}$ eV. In this case,
both $\langle \ln A \rangle$ and $\sigma_{{\ln} A}$ vary with energy.
\end{enumerate}

%%%%%%%%%%%%%%%%%%%%%%%%%%%%%%%%%%%%%%%%%%%%%%%%%%%%%%%%%%%%%%%%%%%%%%%%%%%%
\begin{figure}[!htb]
\centering
\begin{tabular}{l}
\includegraphics*[width=.8\textwidth]{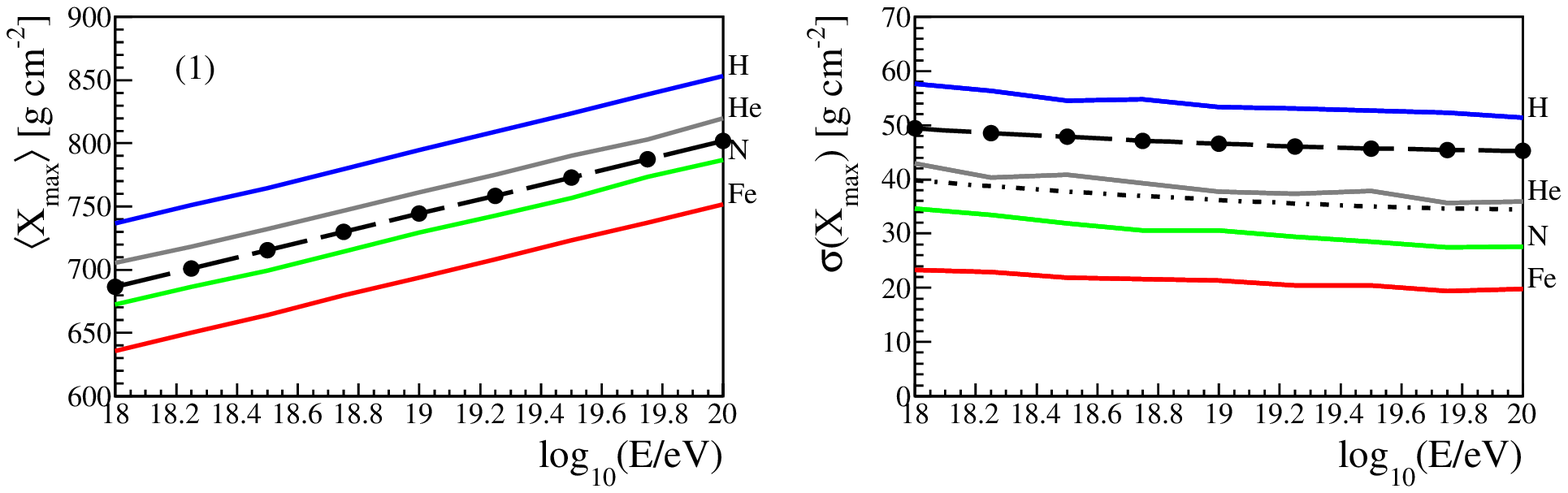}\\
\includegraphics*[width=.8\textwidth]{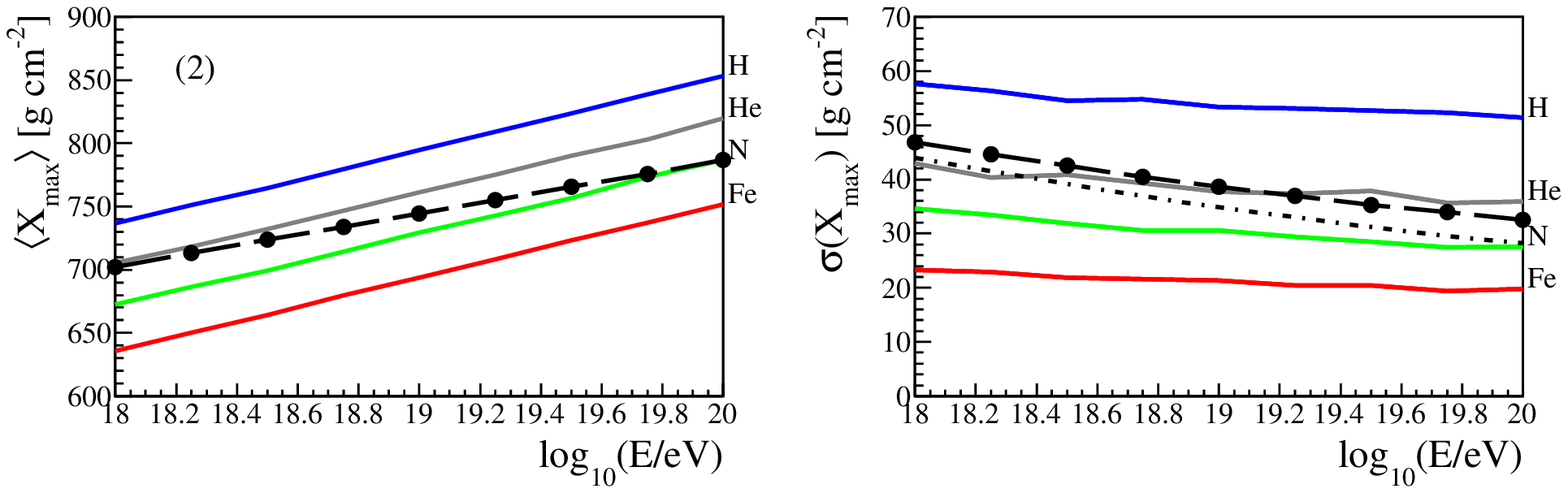}\\
\includegraphics*[width=.8\textwidth]{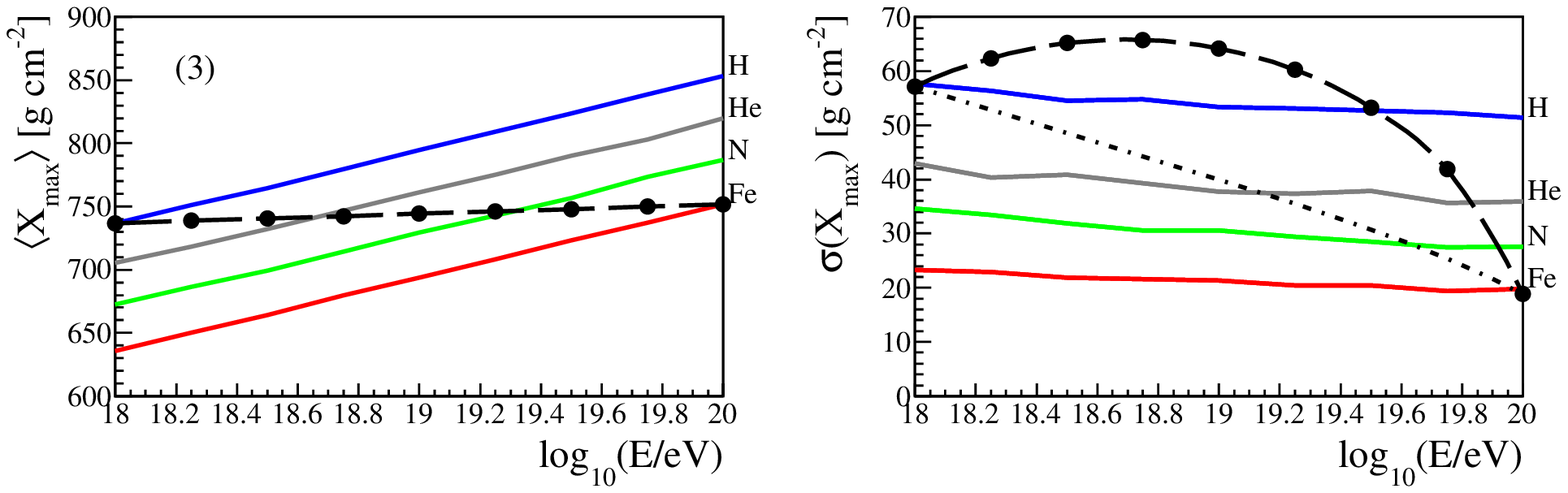}
\end{tabular}
\caption {\label{fig:valid}{\small{$\langle X_\mathrm{max}\rangle$ 
and $\sigma (X_\mathrm{max})$ as a function of $\log_{10}(E/\mathrm{eV})$ for three
different mass distribution hypotheses (see text).
Full circles are calculated from the resulting $X_\mathrm{max}$ distributions from the 
CONEX simulations. Sibyll~2.1 has been chosen for hadronic interactions.  
The dashed lines show equations (\ref{eq:modXmax}) 
for $\langle X_\mathrm{max}\rangle$ 
and (\ref{eq:modSXmax}) for $\sigma (X_\mathrm{max})$. 
The dot-dashed line refers to the 
contribution of the first term in (\ref{eq:modSXmax}). 
 }}}
\end{figure}
%%%%%%%%%%%%%%%%%%%%%%%%%%%%%%%%%%%%%%%%%%%%%%%%%%%%%%%%%%%%%%%%%%%%%%%%%%%%

Figure  \ref{fig:valid} shows the result of the test for the three mass distribution hypotheses.
To generate the $X_\mathrm{max}$ distributions we have used CONEX~\cite{conex} showers with 
Sibyll~2.1~\cite{Sibyll} as the hadronic interaction model. These distributions
do not include detector effects. For each test mass hypothesis, 
the mean and RMS are retrieved from the resulting $X_\mathrm{max}$ distribution obtained 
from the simulations.
These are shown as full circles, $\langle X_\mathrm{max}\rangle$  and $\sigma (X_\mathrm{max})$ in left and right panels respectively. 
The dashed lines are calculated using equations (\ref{eq:modXmax}) and (\ref{eq:modSXmax}) 
for the three different mass hypotheses by using only the first two moments 
$\langle \ln A \rangle$ and $\sigma_{{\ln} A}$. 

One can see that, despite the simple assumptions made,
good agreement is achieved for all 
the three mass distributions. The dot-dashed line refers to the 
contribution of the first term in eq. (\ref{eq:modSXmax}). 
The comparison between the two lines (dashed vs. dot-dashed) highlights how 
different the interpretation of $\sigma (X_\mathrm{max})$ data can be if one 
does not take into account the mass dispersion term.

\begin{figure}[!htb]
\centering
\begin{tabular}{l}
\includegraphics*[width=.8\textwidth]{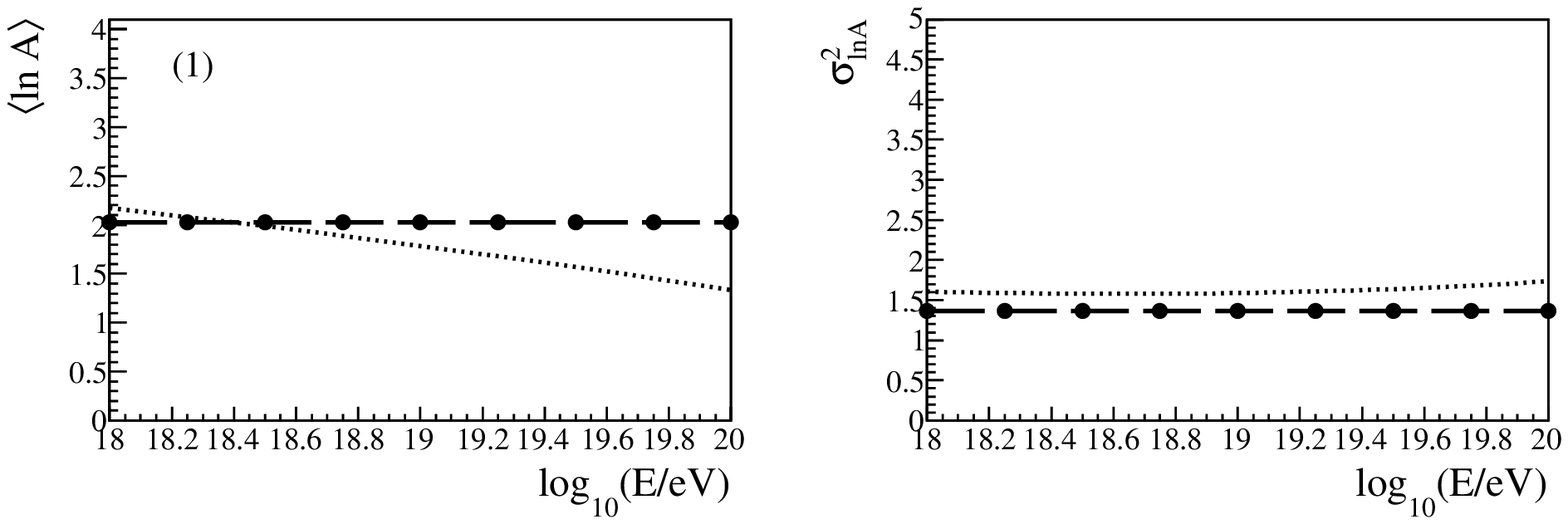}\\
\includegraphics*[width=.8\textwidth]{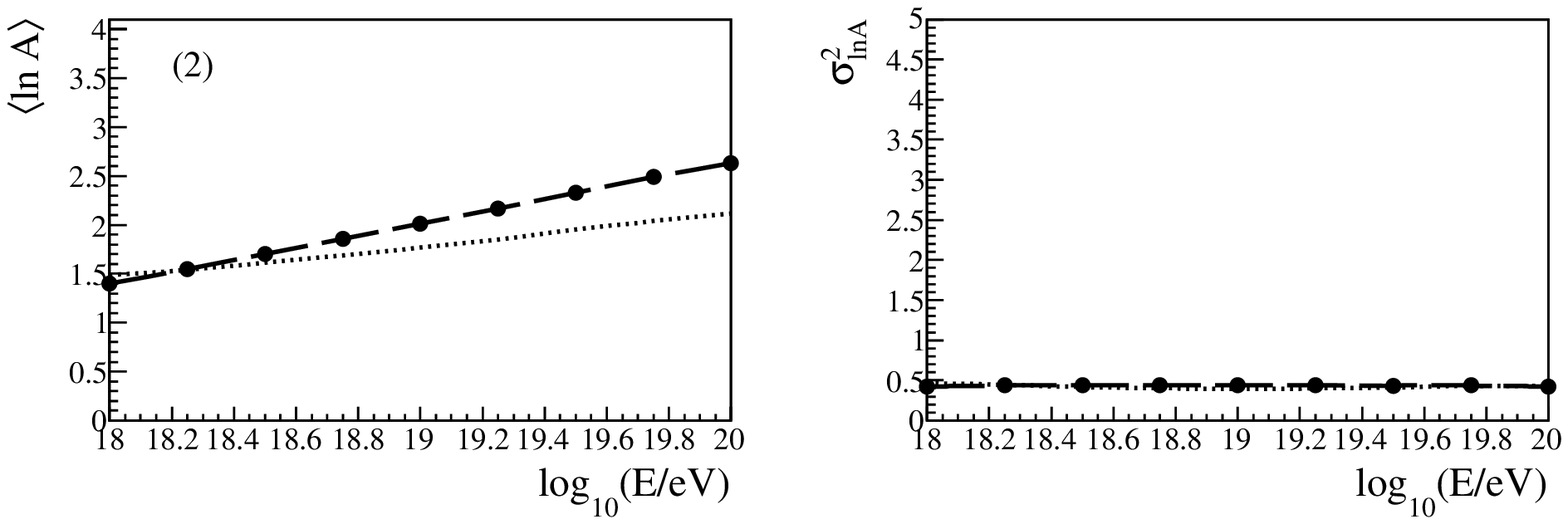}\\
\includegraphics*[width=.8\textwidth]{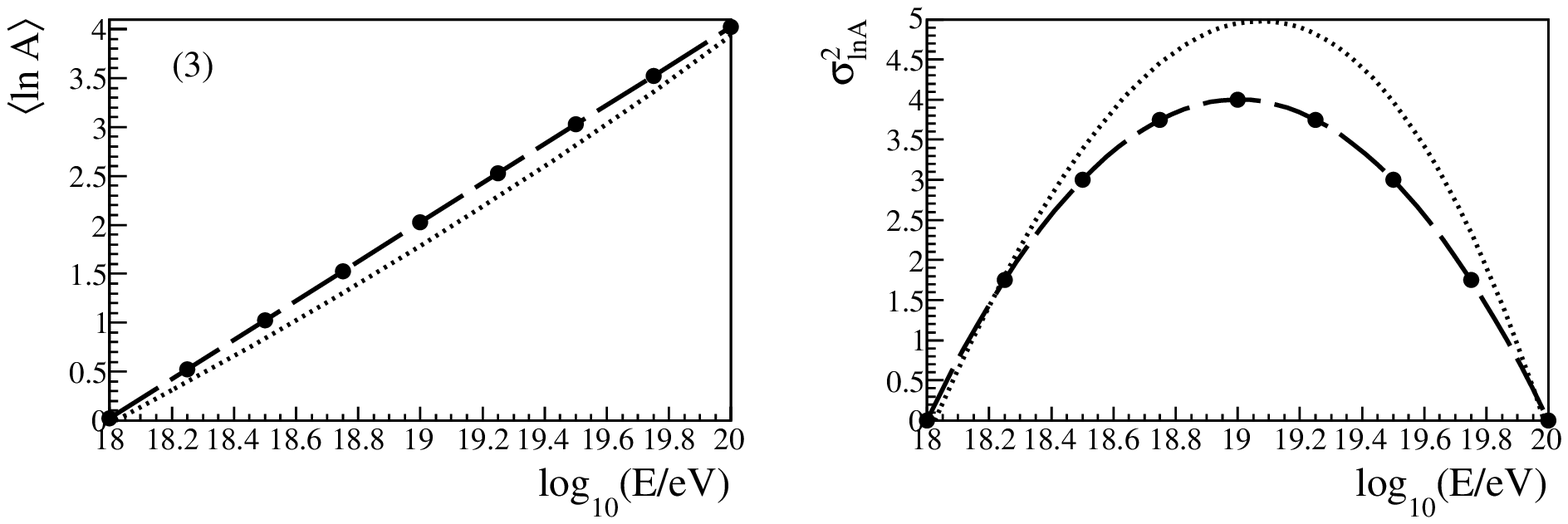}
\end{tabular}
\caption {\label{fig:validInv}{\small{$\langle \ln A \rangle$ 
and $\sigma^2_{\ln A}$ as a function of $\log_{10}(E/\mathrm{eV})$ for three
different mass distribution hypotheses. 
Sibyll~2.1 is the hadronic interaction model.
Full circles refer to the values obtained directly from the input mass
distributions.  The dashed lines show
$\langle \ln A \rangle$ and $\sigma^2_{\ln A}$ calculated using equations
(\ref{eq:meanlnAinv}) and (\ref{eq:VarlnAinv}).
The dotted lines refer to the calculation of the same variables using the
parameterization for QGSJet~II in (\ref{eq:meanlnAinv}) and (\ref{eq:VarlnAinv}).
}}}
\end{figure}

The inverse equations (\ref{eq:meanlnAinv}) and (\ref{eq:VarlnAinv}) have also 
been tested using Monte Carlo simulation. In this case $\langle \ln A \rangle$ 
and $\sigma^2_{\ln A}$ have been obtained as a function of $\log_{10}(E/\mathrm{eV})$ directly from the input mass distributions. 
These values are shown as full circles in Figure \ref{fig:validInv}. 
The $\langle X_\mathrm{max}\rangle$ 
and $\sigma (X_\mathrm{max})$ retrieved from the corresponding $X_\mathrm{max}$ 
distributions are used in equations (\ref{eq:meanlnAinv}) 
and (\ref{eq:VarlnAinv}) to get $\langle \ln A \rangle$ 
and $\sigma^2_{\ln A}$. These are shown in
Fig. \ref{fig:validInv} as dashed lines. Also in this case, the comparison
is quite successful.
%%%%%%%%%%%%%%%%%%%%%%%%%%%%%%%%%%%%%%%%%%%%%%%%%%%%%%%%%%%%%%%%%%%%%%%%%%%%
%%%%%%%%%%%%%%%%%%%%%%%%%%%%%%%%%%%%%%%%%%%%%%%%%%%%%%%%%%%%%%%%%%%%%%%%%%%%

The simulated data sample can also be used to estimate the systematic uncertainty
in the calculation of the moments of the $X_\mathrm{max}$ ($\ln A$) distribution induced by
the missing knowledge of the hadronic interaction mechanism. This study is
pursued using simulated showers generated with a given model together with parameters
of another model in equations (\ref{eq:modXmax}) and (\ref{eq:modSXmax}) for the 
profile variables,
and (\ref{eq:meanlnAinv}) and (\ref{eq:VarlnAinv}) for the log mass variables. An example
of this procedure is shown in Fig. \ref{fig:validInv} where the dotted lines show
the calculation with the parameters of QGSJet~II and the full circles refer to data simulated
with Sibyll 2.1. As a summary of these cross-model checks, we find mean absolute deviations of
4 to 27 g cm$^{-2}$ for $\langle X_\mathrm{max}\rangle$  and 1 to 5.4  g cm$^{-2}$ for
$\sigma (X_\mathrm{max})$, where the maximum deviations are obtained crossing EPOS with QGSjetII. 
The same study done for the moments of the log mass distribution gives 
mean absolute deviations of 0.2 to 1.2 for 
$\langle \ln A \rangle$ and 0.02 to 0.5 for $\sigma^2_{\ln A}$. In this case the maximum values 
refer to EPOS vs. QGSJet 01 for the first moment and QGSJet~II vs. QGSJet~01 for the second.

\section{Application to data  \label{sec.Data}}

At ultra-high energies, shower development can be directly measured 
using fluorescence and Cherenkov light profiles. 
Mean $X_\mathrm{max}$ data as a function of energy are available from 
Fly's Eye~\cite{FlysEye}, HiRes~\cite{HiRes1,HiRes2}, Auger~\cite{Auger}, Yakutsk~\cite{Yakutsk} 
and Telescope Array~\cite{TA}. $\langle X_\mathrm{max}\rangle$ data
were complemented  with fluctuation measurements as early as 1980s 
(see e.g.~\cite{WatsonFluct} and references therein) but only recently have precise optical 
detector measurements become available~\cite{HiRes2,Auger,Yakutsk}.
 
\begin{figure}[!htb]
\centering
\includegraphics*[width=\textwidth]{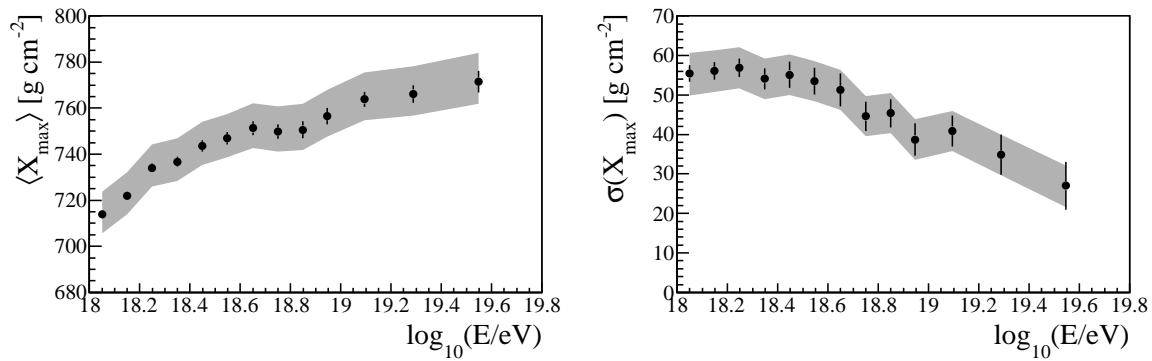}
\caption {\label{fig:XmaxData}{\small{$\langle X_\mathrm{max} \rangle$ (left)
and $\sigma(X_\mathrm{max})$ (right) as a function of $\log_{10}(E/\mathrm{eV})$
from Pierre Auger Observatory data~\cite{AugerICRC2011,AugerTable}.
Data (full circles) are shown with statistical errors. Systematic uncertainties
are represented as bands.
}}}
\end{figure}

The Pierre Auger Collaboration has published results on the mean and 
dispersion of the $X_\mathrm{max}$ distribution at energies above 
10$^{18}$ eV~\cite{Auger}. Here we 
apply the method presented in this work to an updated dataset available 
in ~\cite{AugerICRC2011,AugerTable}. These data are shown in 
Figure~\ref{fig:XmaxData}.
%%%%%%%%%%%%%%%%%%%%%%%%%%%%%%%%%%%%%%%%%%%%%%%%%%%%%%%%%%%%%%%%%%%%%%%%%%%%
%%%%%%%%%%%%%%%%%%%%%%%%%%%%%%%%%%%%%%%%%%%%%%%%%%%%%%%%%%%%%%%%%%%%%%%%%%%%

In the Auger analysis~\cite{Auger}, the events are selected using 
fiducial volume cuts based on the shower geometry. This ensures that the viewable  
$X_\mathrm{max}$  range for each shower is large enough to accommodate the full  
$X_\mathrm{max}$ distribution.
Also, the detector resolution is accounted for by subtracting in quadrature its
contribution to the measured dispersion. 
This allows the direct conversion to the moments of the $\ln A$ distribution 
using equations (\ref{eq:meanlnAinv}) and (\ref{eq:VarlnAinv}) without 
the need of more complex treatment, such as is required in the presence
of  acceptance biases~\cite{UngerICRC2011,Cazon}.

\begin{figure}[!htb]
\centering
\includegraphics*[width=.9\textwidth]{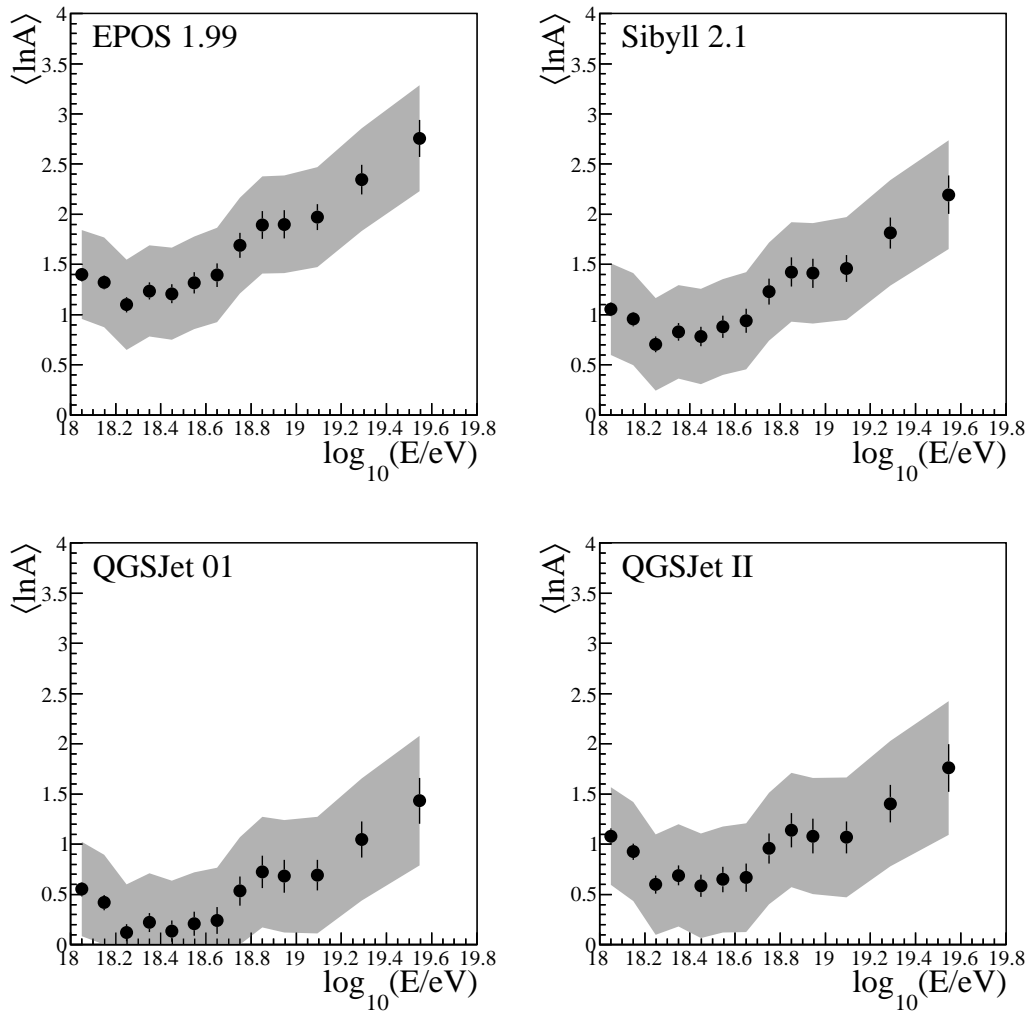}
\caption {\label{fig:meandata}{\small{$\langle \ln A \rangle$ 
as a function of $\log_{10}(E/\mathrm{eV})$  obtained from Auger 
data~\cite{AugerTable} are shown as full circles for different 
hadronic interaction models. 
Error bars show statistical errors. The shaded areas refer to 
systematic uncertainties obtained by summing 
in quadrature the systematic uncertainties 
on $\langle X_\mathrm{max}\rangle$ and 
$\sigma (X_\mathrm{max})$ data points and on the FD energy scale. 
}}}
\end{figure}

\begin{figure}[!htb]
\centering
\includegraphics*[width=.9\textwidth]{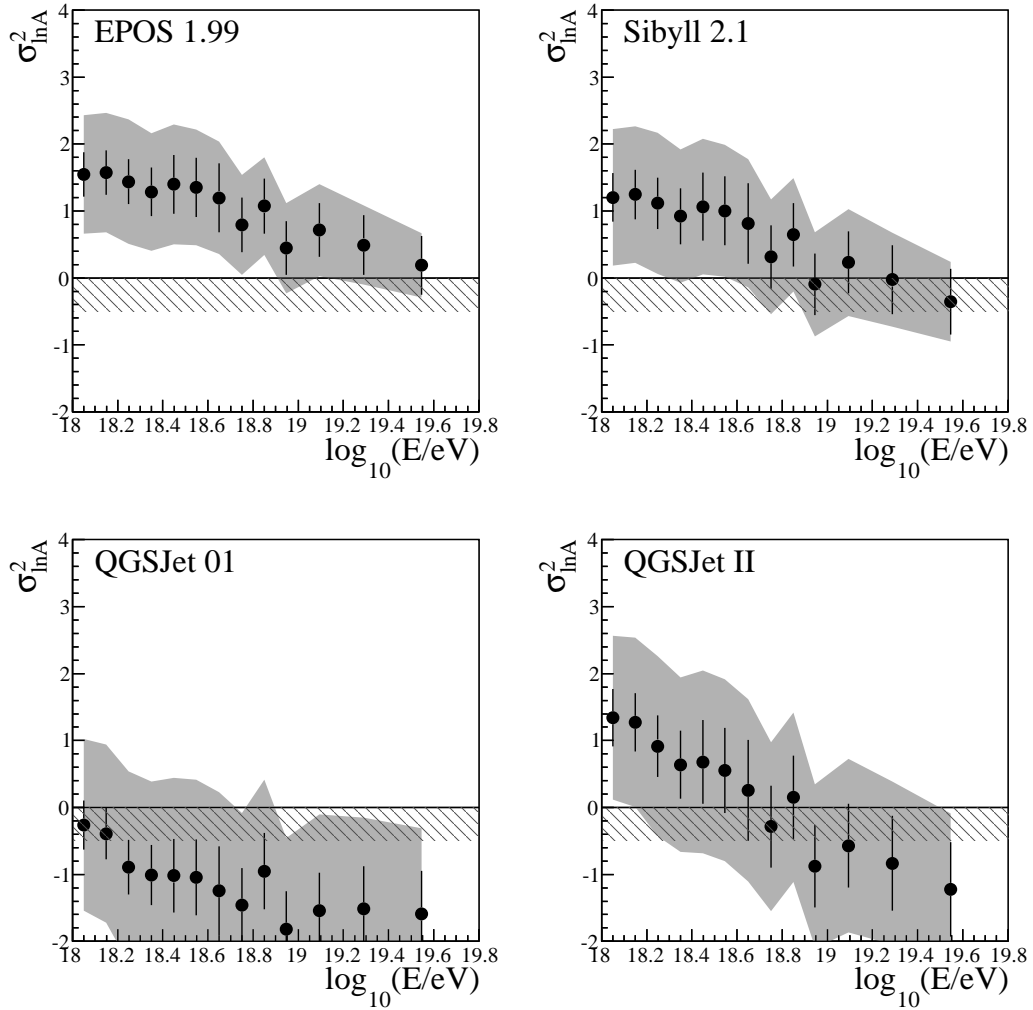}
\caption {\label{fig:vardata}{\small{$\sigma^2_{\ln A}$ as a function of 
$\log_{10}(E/\mathrm{eV})$ obtained from Auger data~\cite{AugerTable} 
are shown as full circles for different  hadronic interaction models. 
Error bars show statistical errors. The shaded area refers to systematic
 uncertainties as in Fig.~\ref{fig:meandata}. The lower limit of
allowed $\sigma^2_{\ln A}$ is shown by the exclusion line. The upper
limit (4.05) is just above the maximum of the vertical axis.
}}}
\end{figure}

The moments of the log mass distribution,
 $\langle \ln A \rangle$ and $\sigma^2_{\ln A}$, 
as obtained using equations 
(\ref{eq:meanlnAinv}) and (\ref{eq:VarlnAinv}), are shown (full circles) 
as a function 
of $\log_{10}(E/\mathrm{eV})$ in 
Figures~\ref{fig:meandata} and \ref{fig:vardata} respectively.
Error bars show the statistical errors obtained from the
propagation of data errors and the errors of the fitted parameters. 
Shaded bands are the systematic uncertainties obtained by summing in 
quadrature the different individual contributions. 
%%%%%%%%%%%%%%%%%%%%%%%%%%%%%%%%%%%%%%%%%%%%%%%%%%%%%%%%%%%%%%%%%%%%%%%%%%%%
%%%%%%%%%%%%%%%%%%%%%%%%%%%%%%%%%%%%%%%%%%%%%%%%%%%%%%%%%%%%%%%%%%%%%%%%%%%%
%%%%%%%%%%%%%%%%%%%%%%%%%%%%%%%%%%%%%%%%%%%%%%%%%%%%%%%%%%%%%%%%%%%%%%%%%%%%
%%%%%%%%%%%%%%%%%%%%%%%%%%%%%%%%%%%%%%%%%%%%%%%%%%%%%%%%%%%%%%%%%%%%%%%%%%%%
The systematic uncertainties on $\langle X_\mathrm{max}\rangle$ and 
$\sigma (X_\mathrm{max})$ data points have different sources: 
calibration, atmospheric conditions, reconstruction and event 
selection~\cite{Auger}.  
Another source of systematics is related to the
uncertainty of the FD energy scale~\cite{EnergyScale}, 22~\%, which 
induces an uncertainty in $\langle \ln A \rangle$ and $\sigma^2_{\ln A}$
 via the parameters of the models. 
All these uncertainties contribute approximately at the same level and 
independently of energy.
The figures show the results for the moments of the log
mass distribution for EPOS 1.99~\cite{epos}, Sibyll~2.1~\cite{Sibyll}, 
QGSJet~01~\cite{qgsjet01} and QGSJet~II~\cite{qgsjetII}.

Despite the uncertainties and the different mass offsets of the models, the overall features are similar in all
the cases. So far as the energy dependence is concerned, the data imply an increasing 
$\langle \ln A \rangle$ 
above 10$^{18.3}$ eV from light to intermediate masses  and 
a decreasing $\sigma^2_{\ln A}$ over the whole energy range.

Looking more specifically to the different hadronic models we notice a slight change 
in the log mass scale. The highest masses are obtained for  EPOS~1.99.  Sibyll~2.1
and QGSJet~II show intermediate values, whereas the lowest masses are obtained for 
QGSJet~01. In particular at $\log_{10}(E/\mathrm{eV}) = 18.25$ the mean log
mass, $\langle \ln A \rangle$, is 1.10, 0.70, 0.60 and 0.12 respectively
for EPOS~1.99, Sibyll~2.1, QGSJet~II and QGSJet~01 with statistical errors of about 0.08
and systematic uncertainty of about 0.6. 
The Pierre Auger Collaboration has recently published
the measurement of the proton-air cross 
section for the energy interval 10$^{18}$ to 10$^{18.5}$ eV~\cite{AugerXsect}. 
That measurement is done using the showers with $X_\mathrm{max}\geq 768$ g cm$^{-2}$, 
corresponding to 20\% of the total $X_\mathrm{max}$ distribution. 
Even in the most unfavourable case, (the $\langle \ln A \rangle$ 
and $\sigma^2_{\ln A}$ predicted by EPOS), one finds that several realizations obtained
from the allowed $\langle \ln A \rangle$ and $\sigma^2_{\ln A}$ have enough protons in the most deeply 
penetrating showers to fulfill the selection criteria adopted in the Auger analysis.

Whereas $\langle \ln A \rangle$ always has valid
values (apart a small region which crosses $\langle \ln A \rangle = 0$ for
QGSJet~01), there are wide energy intervals where $\sigma^2_{\ln A}$ is negative.
Considering eq. (\ref{eq:VarlnAinv})  one can  see
that these values 
occur for energies where the shower fluctuations corresponding to the mean 
log mass exceed the measured $X_\mathrm{max}$ fluctuations. 
Figure~\ref{fig:vardata} shows that 
$\sigma^2_{\ln A}$ data points are within the allowed physical region
only for EPOS 1.99 and Sibyll~2.1. They are partly outside 
for QGSJet~II, and completely outside for QGSJet~01. However the current
systematic uncertainties do not allow one to establish stringent tests
to the models. 

The method presented in this work shows that the Pierre Auger Observatory 
data can confront 
hadronic physics models provided that future developments in the shower data
analysis reduce systematics. By shrinking the shaded bands in 
Figure~\ref{fig:vardata}  it will be possible to constrain those models.

\section{Discussion  \label{sec.discussion}}
The importance of the combined study of the mean values and fluctuations of mass
dependent observables has been addressed by 
several authors~\cite{WatsonFluct,Watson1,Watson2,Linsley83,Linsley85}. In particular, 
Linsley~\cite{Linsley85} showed that a combined analysis of the mean and the
variance of $\ln A$ can provide a useful representation of the mass transition 
(if any) to be found in shower profile data. In fact, possible transitions 
are constrained to a limited region of the ($\langle \ln A \rangle$, $\sigma^2_{\ln A}$) plane. More recently 
a similar study using the 
$\langle X_\mathrm{max} \rangle$-$\sigma(X_\mathrm{max})$ correlation\footnote{
In this case
the dependence on hadronic models has been accounted for by
subtracting the corresponding
observables predicted by the models for iron.}
reached a similar conclusion~\cite{KHK-MU-review}.

\begin{figure}[!htb]
\centering
\includegraphics*[width=.9\textwidth]{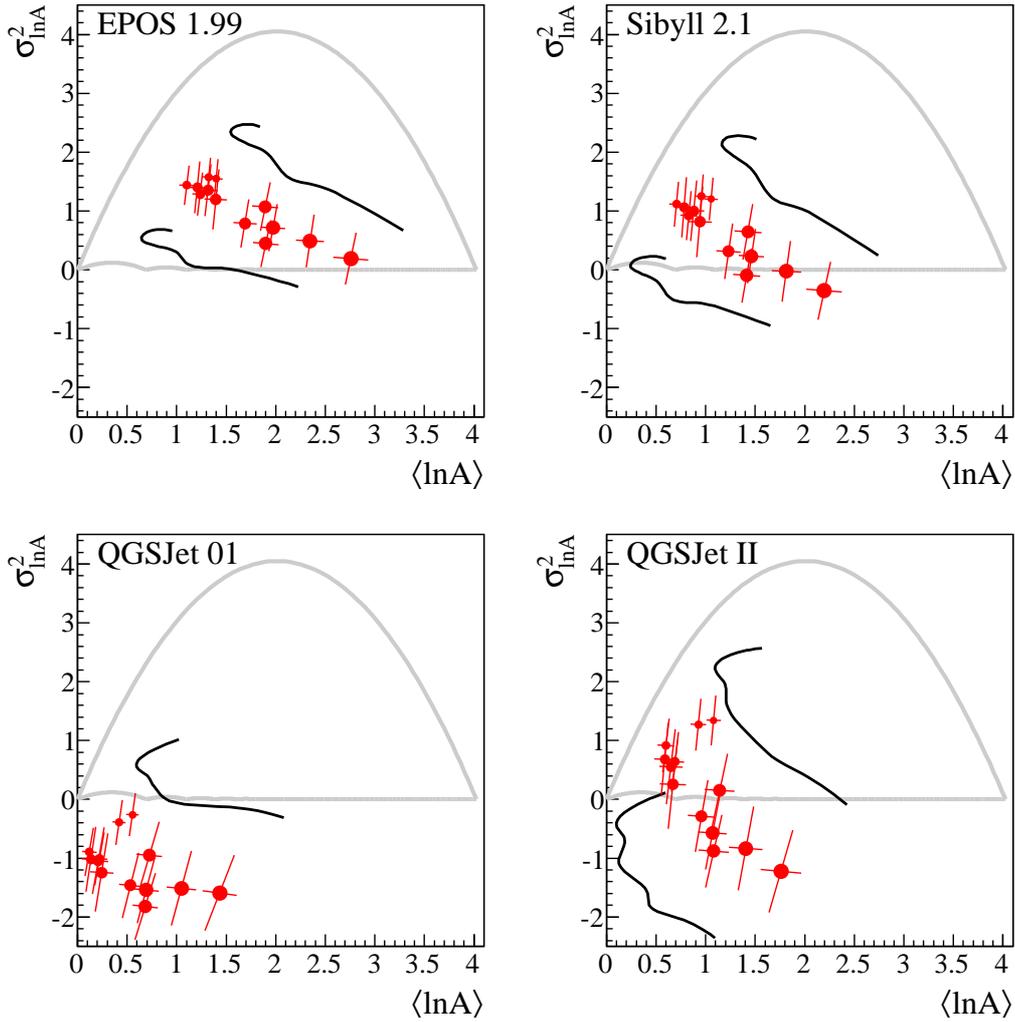}
\caption {\label{fig:umbrella}{\small{Pierre Auger data in the 
($\langle \ln A \rangle$, $\sigma^2_{\ln A}$) plane for different hadronic 
interaction models. Data points are shown as full circles with statistical errors. The marker
sizes increase with the logarithm of the energy. Systematic 
uncertainties are shown as solid lines.
The gray thick line shows the contour of the
$\langle \ln A \rangle$ and $\sigma^2_{\ln A}$ values allowed 
for nuclear compositions. }}}
\end{figure}

Converting $X_\mathrm{max}$ data to $\ln A$ variables, as described in 
Sec.~\ref{sec.Stat}, one can plot Pierre Auger Observatory data in the 
($\langle \ln A \rangle$, $\sigma^2_{\ln A}$) plane. Since this procedure 
depends on the hadronic model, one gets a plot for each model as shown in
Figure~\ref{fig:umbrella}. Data points  are shown as full circles with
size increasing in proportion to $\log E$. The error bars are tilted because 
of correlations
arising from equations (\ref{eq:meanlnAinv}) and (\ref{eq:VarlnAinv})
and represent the principal axes of the statistical error ellipses.
The solid lines show the systematic uncertainties.
The same figure shows the region allowed for mass compositions. The contour
of this region  (gray thick line) is generated by mixing neighbouring nuclei in 
the lower edge and extreme nuclei (protons and iron) in the upper 
edge. Each of these mixings is an arch shaped line in the  
($\langle \ln A \rangle$, $\sigma^2_{\ln A}$) plane.

%%%%%%%%%%%%%%%%%%%%%%%%%%%%%%%%%%%%%%%%%%%%%%%%%%%%%%%%%%%%%%%%%%%%%%%%%%%%
%%%%%%%%%%%%%%%%%%%%%%%%%%%%%%%%%%%%%%%%%%%%%%%%%%%%%%%%%%%%%%%%%%%%%%%%%%%%

Figure \ref{fig:umbrella} shows that the Auger data lie outside the 
allowed boundaries for part of the energy range in some of the models. 
As noted previously, systematic uncertainties are still large and thus prevent us from
more definite conclusions. 
However the energy evolution is common to all models 
suggesting that the average mass increases with decreasing log mass dispersion.
This behaviour might imply astrophysical consequences.

In fact there are only a few possibilities for extragalactic source
models to  produce compositions with small log mass dispersion at the Earth.
Protons can traverse their path from sources to the Earth without
mass dispersion, but this case is excluded  by Pierre Auger Observatory 
data at the highest energies.

Nuclei originating from nearby sources ($\lessapprox$ 100 Mpc) 
might be detected with small mass dispersion. 
For these sources, propagation does not degrade mass and energy 
so the spectrum and composition reflect closely their values at  injection. 
But, if sources are distributed uniformly, distant sources induce
natural mass dispersions. 
Small $\ln A$ dispersions are possible only when there is small observed mass 
mixing so that, at each energy, only nuclei with a small spread in masses are present. This
corresponds to the low-$\sigma^2_{\ln A}$ edge of the contour of the
allowed region in the ($\langle \ln A \rangle$, $\sigma^2_{\ln A}$) plane.
 
Protons originating by the 
photo-disintegration of nuclei are the main source of mass dispersion 
because they populate each energy region. The possible end of the injection
spectrum based on a rigidity-dependent mechanism can reduce the proton component
at high energies, thus producing a reduction of the mass dispersion at
the highest energies.
A complete study of source models under several hypotheses is required
to study all the source parameters that limit the mass dispersion
in the propagation of extragalactic cosmic rays. 
Recent studies, see e.g.~\cite{Taylor,Denis}, based on the assumption of a
uniform source distribution,
have shown that the Auger composition results, when combined
with the energy spectrum, require hard injection spectra (i.e. index \textless ~2)
 with low energy cutoffs and the possible presence of local sources.

\section{Conclusions   \label{sec.Conc}}
In this work we presented a method for interpreting 
$\langle X_\mathrm{max} \rangle$ 
and $\sigma (X_\mathrm{max})$ in terms of mass composition. The method is 
 based on an extension of the Heitler model of extensive air showers. 
The parameterization given in equations (\ref{eq:modXmax}) and 
(\ref{eq:modSXmax}) expresses those
two profile observables as a linear combination of the first two moments of 
the log mass distribution, $\langle \ln A \rangle$ and $\sigma^2_{\ln A}$, and of 
the mean shower fluctuations. 

We first note that the method provides an effective key to the interpretation of data. The energy dependences of
$\langle X_\mathrm{max} \rangle$ and
$\sigma(X_\mathrm{max})$ are sometimes considered as different
expressions of the same 
physical features, e.g. an increase or decrease of the mean log mass. However 
their different meanings can be easily understood by looking at the 
dependence on the mass variables.
At a fixed energy $\langle X_\mathrm{max} \rangle$ is only function
of $\langle \ln A \rangle$; therefore, it only carries information of the average composition. However, $\sigma(X_\mathrm{max})$ 
cannot be interpreted as a measure of the average composition since it is also
affected by the log mass dispersion. 
Similarly, the inference of hadronic interaction properties from $\sigma(X_\mathrm{max})$ can be wrong unless the mass 
dispersion term ($\propto \sigma^2_{{\ln} A}$) is negligible. 
The parameter $\sigma_{{\ln} A}$ represents the dispersion of the masses as they hit the Earth atmosphere. It reflects not only the
spread of nuclear masses at the sources but also the
modifications that occur during their propagation to the
Earth.

The method has been succesfully tested, with the simulation of different mass 
distributions in the energy interval from 10$^{18}$ to 10$^{20}$ eV showing the 
robustness of the parameterization. 
We have applied the method to the Pierre Auger Observatory $X_\mathrm{max}$
data to get the first two moments of the $\ln A$ distribution.
The outcome relies on the choice of
a hadronic interaction model to set the parameters and the appropriate
shower fluctuations. Four models have been used,  EPOS 1.99, Sibyll~2.1, 
QGSJet~01 and QGSJet~II, and the corresponding  moments of the log mass 
distribution have been obtained as a function of energy. 
Despite the differences in the chosen models, the overall features are quite 
similar. In particular we find an increasing $\langle \ln A \rangle$ 
above 10$^{18.3}$ eV from light to intermediate masses  and 
a decreasing $\sigma^2_{\ln A}$ over the whole energy range, while the mean log mass scale
changes with hadronic models. 

The results presented in this paper show the capability of the method to infer 
important features
of the mass distribution of the UHECR nuclei.
This is a remarkable outcome with respect to the study of source 
scenarios and propagation. 
In fact we do not only access the average mass, but also the mass dispersion. 
While a pure proton beam at the sources is not changed by propagation, nuclei 
should increase the mass dispersion in their path towards the Earth. 
The Auger results seem to imply either close-by sources or hard 
spectral indices, if the energy evolution of the present hadronic interaction 
models can be trusted.

The proposed method can also be used as
a tool to investigate the validity of hadronic interaction
models. In particular it has been shown that the intrinsic shower 
fluctuations are sometimes
larger than the measured $X_\mathrm{max}$ dispersions. This happens in different 
energy intervals for the different models. 
At the highest energies, all models approach the lower boundary, 
and some of them enter the unphysical region, 
but the current systematic uncertainties prevent us from confidently rejecting 
any model. Provided that systematic uncertainties can be reduced in 
future data analysis, the method can be used to constrain
hadronic interaction models.
The addition of new measurements, such as  the muon content of 
EAS~\cite{AugerMuons1,AugerMuons2}, may allow us to place stronger bounds to the models.

\appendix
\section{Parameterization of the shower mean depth and its fluctuations  \label{sec.Pars}}
The shower code chosen for this work is CONEX~\cite{conex}. CONEX is a 
hybrid simulation code that is suited for fast
one-dimensional simulations of shower profiles, including  fluctuations.
It combines Monte Carlo simulation of high energy interactions with
a fast numerical solution of cascade equations for the resulting
distributions of secondary particles. In our CONEX simulation we 
used the default energy thresholds settings of version v2r3.1\footnote{
The hadron and muon cutoff (minimum energy) is 1 GeV, the 
cutoff for electrons, positrons and gammas (e/m particles) is 1 MeV, the threshold energy
for solving cascade equations is 0.05, 0.0005 and 0.005 of the primary energy for hadrons,
muons and e/m particles respectively. The above-threshold e/m interaction are simulated with
the EGS4 program~\cite{EGS4}. The low energy (E $<$ 80 GeV) interaction model is 
GEISHA~\cite{GEISHArep}
}
. 

The parameters $X_{0}$, $D_{1}$, $\xi$ and $\delta$ used in equations (\ref{eq:modXmax}) 
and (\ref{eq:modSXmax}) 
have been obtained by fitting CONEX showers for four different primaries 
(H, He, N and Fe) in nine energy bins of width $\Delta \log_{10} (E/\mathrm{eV})$ = 0.25
ranging from 10$^{18}$ to 10$^{20}$ eV, and for all the hadronic models used in 
this work: 
EPOS 1.99~\cite{epos}, Sibyll~2.1~\cite{Sibyll}, QGSJet~01~\cite{qgsjet01}
and QGSJet~II~\cite{qgsjetII}. In total, about 25,000 showers have been used for each
energy bin and for each hadronic model. The fit procedure always converges 
with mean (maximum) $\langle X_\mathrm{max} \rangle$ residuals from the
simulated data  of about 1 (3) g cm$^{-2}$ for all the models.
The best fit values are reported in  Table~\ref{tab:modpar} with their errors.

%%%%%%%%%%%%%%%%%%%%%%%%%%%%%%%%%%%%%%%%%%%%%%%%%%%%%%%%%%%%%%%%%%%%%%%%%%%
\begin{table}[!htb]
\centering
\begin{tabular}{|c||c|c|c|c|}
\hline
parameter &EPOS 1.99 & Sibyll 2.1 & QGSJet 01 & QGSJet II \\
\hline
\hline
$X_{0}$ & 809.7 $\pm$ 0.3 & 795.1 $\pm$ 0.3 & 774.2 $\pm$ 0.3 & 781.8 $\pm$ 0.3\\
\hline

$D$ & 62.2 $\pm$ 0.5 & 57.7 $\pm$ 0.5  & 49.7 $\pm$ 0.5 & 45.8 $\pm$ 0.5\\
\hline

$\xi$ &  0.78 $\pm$ 0.24 & -0.04 $\pm$ 0.24 & -0.30 $\pm$ 0.24 & -1.13 $\pm$  0.24\\
\hline

$\delta$  &  0.08 $\pm$ 0.21 &  -0.04 $\pm$ 0.21 & 1.92 $\pm$ 0.21 & 1.71 $\pm$ 0.21\\

\hline
\hline
\end{tabular}
\caption{\label{tab:modpar}{\small Parameters of formulae (\ref{eq:modXmax}) and (\ref{eq:modSXmax}) for different hadronic interaction models setting $E_0$ = 10$^{19}$ eV. The values are obtained fitting the mean $X_\mathrm{max}$ for showers generated for four different primaries H, He, N and Fe. 
Statistical error obtained from the fit are also given. All values are expressed
 in g cm$^{-2}$.}}
\end{table}
%%%%%%%%%%%%%%%%%%%%%%%%%%%%%%%%%%%%%%%%%%%%%%%%%%%%%%%%%%%%%%%%%%%%%%%%%%%%

Shower variances have been fitted using the parameterization given
in equations (\ref{eq:sigsh}) and (\ref{eq:sigshE}) and the same simulated data set
described above. The mean (maximum) $\sigma(X_\mathrm{max})$ residuals from the
simulated data are about 1 (3) g cm$^{-2}$ for all the models.
The best fit parameters are given in Table~\ref{tab:varsh} with their errors.
%%%%%%%%%%%%%%%%%%%%%%%%%%%%%%%%%%%%%%%%%%%%%%%%%%%%%%%%%%%%%%%%%%%%%%%%%%%
\begin{table}[!htb]
\centering
\begin{tabular}{|c||c|c|c|c|}
\hline
parameter &EPOS 1.99 & Sibyll 2.1 & QGSJet 01 & QGSJet II \\
\hline
\hline
$p_0 \times (\rm g^{-2} cm^{4})$ & 3279 $\pm$ 51 & 2785 $\pm$  46 &  3852 $\pm$  55 & 3163 $\pm$  49\\
\hline

$p_1 \times (\rm g^{-2} cm^{4})$ & -47 $\pm$  66   &  -364 $\pm$  58 &  -274 $\pm$  70 &  -237 $\pm$    61\\
\hline

$p_2 \times (\rm g^{-2} cm^{4})$ & 228 $\pm$ 108 &  152 $\pm$    93 &  169 $\pm$   116 & 60 $\pm$   100 \\
\hline

$a_0$  & -0.461 $\pm$ 0.006  &  -0.368 $\pm$ 0.008 & -0.451 $\pm$ 0.006 & -0.386 $\pm$ 0.007\\
\hline

$a_1$  & -0.0041 $\pm$ 0.0016 &  -0.0049 $\pm$ 0.0023 &  -0.0020 $\pm$ 0.0016 & -0.0006 $\pm$ 0.0021\\
\hline

$b$  &  0.059 $\pm$ 0.002 &  0.039 $\pm$  0.002 &  0.057 $\pm$  0.001 & 0.043 $\pm$  0.002\\
\hline
\hline
\end{tabular}
\caption{\label{tab:varsh}{\small Parameters of formulae (\ref{eq:sigsh}) and
(\ref{eq:sigshE}) for different hadronic interaction models setting $E_0$ = 10$^{19}$ eV. The values are obtained fitting $\sigma^2(X_\mathrm{max})$ for showers generated for four different primaries H, He, N and Fe. The statistical errors obtained from the fit are also given.}}
\end{table}
%%%%%%%%%%%%%%%%%%%%%%%%%%%%%%%%%%%%%%%%%%%%%%%%%%%%%%%%%%%%%%%%%%%%%%%%%%%%

\acknowledgments
The successful installation, commissioning, and operation of the Pierre Auger Observatory
would not have been possible without the strong commitment and effort
from the technical and administrative staff in Malarg\"ue.

We are very grateful to the following agencies and organizations for financial support: 
Comisi\'on Nacional de Energ\'ia At\'omica, 
Fundaci\'on Antorchas,
Gobierno De La Provincia de Mendoza, 
Municipalidad de Malarg\"ue,
NDM Holdings and Valle Las Le\~nas, in gratitude for their continuing
cooperation over land access, Argentina; 
the Australian Research Council;
Conselho Nacional de Desenvolvimento Cient\'ifico e Tecnol\'ogico (CNPq),
Financiadora de Estudos e Projetos (FINEP),
Funda\c{c}\~ao de Amparo \`a Pesquisa do Estado de Rio de Janeiro (FAPERJ),
Funda\c{c}\~ao de Amparo \`a Pesquisa do Estado de S\~ao Paulo (FAPESP),
Minist\'erio de Ci\^{e}ncia e Tecnologia (MCT), Brazil;
AVCR AV0Z10100502 and AV0Z10100522, GAAV KJB100100904, MSMT-CR LA08016,
LG11044, MEB111003, MSM0021620859, LA08015, TACR TA01010517 and GA UK 119810, Czech Republic;
Centre de Calcul IN2P3/CNRS, 
Centre National de la Recherche Scientifique (CNRS),
Conseil R\'egional Ile-de-France,
D\'epartement  Physique Nucl\'eaire et Corpusculaire (PNC-IN2P3/CNRS),
D\'epartement Sciences de l'Univers (SDU-INSU/CNRS), France;
Bundesministerium f\"ur Bildung und Forschung (BMBF),
Deutsche Forschungsgemeinschaft (DFG),
Finanzministerium Baden-W\"urttemberg,
Helmholtz-Gemeinschaft Deutscher Forschungszentren (HGF),
Ministerium f\"ur Wissenschaft und Forschung, Nordrhein-Westfalen,
Ministerium f\"ur Wissenschaft, Forschung und Kunst, Baden-W\"urttemberg, Germany; 
Istituto Nazionale di Fisica Nucleare (INFN),
Ministero dell'Istruzione, dell'Universit\`a e della Ricerca (MIUR), Italy;
Consejo Nacional de Ciencia y Tecnolog\'ia (CONACYT), Mexico;
Ministerie van Onderwijs, Cultuur en Wetenschap,
Nederlandse Organisatie voor Wetenschappelijk Onderzoek (NWO),
Stichting voor Fundamenteel Onderzoek der Materie (FOM), Netherlands;
Ministry of Science and Higher Education,
Grant Nos. N N202 200239 and N N202 207238, Poland;
Portuguese national funds and FEDER funds within COMPETE - Programa Operacional Factores de Competitividade through 
Funda\c{c}\~ao para a Ci\^{e}ncia e a Tecnologia, Portugal;
Romanian Authority for Scientific Research ANCS, 
CNDI-UEFISCDI partnership projects nr.20/2012 and nr.194/2012, 
project nr.1/ASPERA2/2012 ERA-NET and PN-II-RU-PD-2011-3-0145-17, Romania; 
Ministry for Higher Education, Science, and Technology,
Slovenian Research Agency, Slovenia;
Comunidad de Madrid, 
FEDER funds, 
Ministerio de Ciencia e Innovaci\'on and Consolider-Ingenio 2010 (CPAN),
Xunta de Galicia, Spain;
The Leverhulme Foundation,
Science and Technology Facilities Council, United Kingdom;
Department of Energy, Contract Nos. DE-AC02-07CH11359, DE-FR02-04ER41300, DE-FG02-99ER41107,
National Science Foundation, Grant No. 0450696,
The Grainger Foundation USA; 
NAFOSTED, Vietnam;
Marie Curie-IRSES/EPLANET, European Particle Physics Latin American Network, 
European Union 7th Framework Program, Grant No. PIRSES-2009-GA-246806; 
and UNESCO.

%%%%%%%%%%%%%%%%%%%%%%%%%%%%%%%%%%%%%%%%%%%%%%%%%%%%%%%%%%%%%%%%%%%%%%%%%%%%

\end{document}